# Preserving privacy for secure and outsourcing for Linear Programming in cloud computing

G.Vidhisha, C.Surekha, S.Sanjeeva Rayudu, U.Seshadri
*Computer Science Engineering*
Jawaharlal Nehru Technological University *Ananatapur*

*Abstract—* Cloud computing is the long dreamed vision of computing as a utility, where users can remotely store their data into the cloud so as to enjoy the on-demand high quality applications and services from a shared pool of configurable computing resources. By data outsourcing, users can be relieved from the burden of local data storage and maintenance. we utilize the public key based homomorphism authenticator and uniquely integrate it with random mask technique to achieve a privacy-preserving public auditing system for cloud data storage security while keeping all above requirements in mind. To support efficient handling of multiple auditing tasks, we further explore the technique of bilinear aggregate signature to extend our main result into a multi-user setting, where TPA can perform multiple auditing tasks simultaneously along with investigates secure outsourcing of widely applicable linear programming (LP) computations. In order to achieve practical efficiency, our mechanism design explicitly decomposes the LP computation outsourcing into public LP solvers running on the cloud and private LP parameters owned by the customer Extensive security and performance analysis shows the proposed schemes are provably secure and highly efficient.

## 1. INTRODUCTION

Cloud Computing has been envisioned as the next-generation architecture of IT enterprise, due to its long list of unprecedented advantages in the IT history: on-demand self-service, ubiquitous network access, location independent resource pooling, rapid resource elasticity, usage-based pricing and transference of risk [1]. As a disruptive technology with profound implications, Cloud Computing is transforming the very nature of how businesses use information technology. One fundamental aspect of this paradigm shifting is that data is being centralized or outsourced into the Cloud. From users' perspective, including both individuals and IT enterprises, storing data remotely into the cloud in a flexible on-demand manner brings appealing benefits: relief of the burden for storage management, universal data access with independent geographical locations, and avoidance of capital expenditure on hardware, software, and personnel maintenances, etc [2]. One fundamental advantage of the cloud paradigm is computation outsourcing, where the computational power of cloud customers is no longer limited by their resource-constraint devices.

By outsourcing the workloads into the cloud, customers could enjoy the literally unlimited computing resources in a pay-per-use manner without committing any large capital outlays in the purchase of hardware and software and/or the operational overhead there in. Despite the tremendous benefits, outsourcing computation to the commercial public cloud is also depriving customers' direct control over the systems that consume and produce their data during the computation, which inevitably brings in new security concerns and challenges towards this promising computing model [3]. On the one hand, the outsourced computation workloads often contain sensitive information, such as the business financial records, proprietary research data, or personally identifiable health information etc. To combat against unauthorized information leakage, sensitive data have to be encrypted before outsourcing [4] so as to provide end to end data confidentiality assurance in the cloud and beyond. However, ordinary data encryption techniques in essence prevent cloud from performing any meaningful operation of the Underlying plaintext data [5],

Examples include cloud service providers, for monetary reasons, reclaiming storage by discarding data that has not been or is rarely accessed, or even hiding data loss incidents so as to maintain a reputation [6]. In short, although outsourcing data into the cloud is economically attractive for the cost and complexity of long-term large-scale data storage, it does not offer any guarantee on data integrity and availability. This problem, if not properly addressed, may impede the successful deployment of the cloud architecture. As users no longer physically possess the storage of their data, traditional cryptographic primitives for the purpose of data security protection can not be directly adopted. Thus, how to efficiently verify the correctness of outsourced cloud data without the local copy of data files becomes a big challenge for data storage security in Cloud Computing. Note that simply downloading the data for its integrity verification is not a practical solution due to the expensiveness in I/O cost and transmitting the file across the network.

Besides, it is often insufficient to detect the data corruption when accessing the data, as it might be too late for recover the data loss or damage. Considering the large size of the outsourced data and the user's constrained resource capability, the ability to audit the correctness of the data in a cloud environment can be formidable and expensive for the cloud users [7]. Therefore, to fully ensure the data security and save the cloud users computation resources, it is of critical importance to enable public audit ability for cloud data storage so that the users may resort to a third party auditor (TPA), who has expertise and capabilities that the users do not, to audit the outsourced data when needed. Based on the audit result, TPA could release an audit report, which would not only help users to evaluate the risk of their subscribed cloud data services, but also be beneficial for the cloud service provider to improve their cloud based service platform [8]. In a word, enabling public risk auditing protocols will play an important role for this nascent cloud economy to become fully established; where users will need ways to assess risk and gain trust in Cloud. Recently, the notion of public audit ability has been proposed in the context of ensuring remotely stored data integrity under different systems and security models [9]. Public audit ability allows an external party, in addition to the user himself, to verify the correctness of remotely stored data. However, most of these schemes [10] do not support the privacy protection of users' data against external auditors, i.e., they may potentially reveal user data information to the auditors. This drawback greatly affects the security of these protocols in Cloud Computing. From the perspective of protecting data privacy, the users, who own the data and rely on TPA just for the storage security of their data, do not want this auditing process introducing new vulnerabilities of unauthorized information leakage towards their data security [11]. Moreover, there are legal regulations, such as the US Health Insurance Portability and Accountability Act (HIPAA) , further demanding the outsourced data not to be leaked to external parties [8]. Exploiting data encryption before outsourcing [12] is one way to mitigate this privacy concern, but it is only complementary to the privacy-preserving public auditing scheme to be proposed in this paper. Without a properly designed auditing protocol, encryption itself can not prevent data from "flowing away" towards external parties during the auditing process. Thus, it does not completely solve the problem of protecting data privacy but just reduces it to the one of managing the encryption keys. Unauthorized data leakage still remains a problem due to the potential exposure of encryption keys.

Specifically, we first formulate private data owned by the customer for LP problem as a set of matrices and vectors. This higher level representation allows us to apply a set of efficient Privacy-preserving problem transformation techniques, including matrix multiplication and affine mapping, to transform the original LP problem into some arbitrary one while protecting the sensitive input/output information. One crucial benefit of this higher level problem transformation method is that existing algorithms and tools for LP solvers can be directly reused by the cloud server. Although the generic mechanism defined at circuit level, e.g. [13], can even allow the customer to hide the fact that the outsourced computation is LP, we believe imposing this more stringent security measure than necessary would greatly affect the efficiency. To validate the computation result, we utilize the fact that the result is from cloud server solving the transformed LP problem. In particular, we explore the fundamental duality theorem together with the piece-wise construction of auxiliary LP problem to derive a set of necessary and sufficient conditions that the correct Customer LP problem Φ.

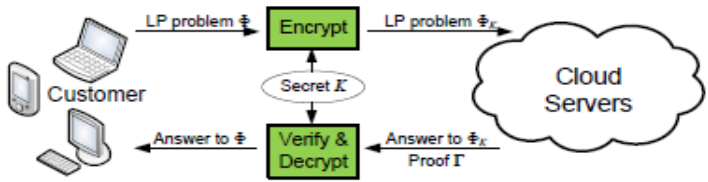

Fig. 1: Architecture of secure outsourcing linear programming problems in Cloud Computing

Result must satisfy. Such a method of result validation can be very efficient and incurs close-to-zero additional overhead on both customer and cloud server. With correctly verified result,

Customer can use the secret transformation to map back the desired solution for his original LP problem. We summarize our contributions as follows:

1) For the first time, we formalize the problem of securely outsourcing LP computations, and provide such a secure and practical mechanism design which fulfills input/output privacy, cheating resilience, and efficiency.
2) Our mechanism brings cloud customer great computation savings from secure LP outsourcing as it only incurs $O(n\_)$ for some $2 < \_ \leq 3$ local computation overhead on the customer, while solving a normal LP problem usually requires more than $O(n3)$ time [13].
3) The computations done by the cloud server shares the same time complexity of currently practical algorithms for solving the linear programming problems, which ensures that the use of cloud is economically viable.
4) The experiment evaluation further demonstrates the immediate practicality: our mechanism can always help customers achieve more than 30× savings when the sizes of the original LP problems are not too small, while introducing no substantial overhead on the cloud.

Therefore, how to enable a privacy-preserving third-party auditing protocol, independent to data encryption, is the problem we are going to tackle in this paper. Our work is among the first few ones to support privacy-preserving public auditing in Cloud Computing, with a focus on data storage. Besides, with the prevalence of Cloud Computing, a foreseeable increase of auditing tasks from different users may be delegated to TPA. As the individual auditing of these growing tasks can be tedious and cumbersome, a natural demand is then how to enable TPA to efficiently perform the multiple auditing tasks in a batch manner, i.e., simultaneously. To address these problems, our work utilizes the technique of public key based homomorphism authenticator [14], which enables TPA to perform the auditing without demanding the local copy of data and thus drastically reduces the communication and computation overhead as compared to the straightforward data auditing approaches. By integrating the homomorphism authenticator with random mask technique, our protocol guarantees that TPA could not learn any knowledge about the data content stored in the cloud server during the efficient auditing process.

## 2. PROBLEM STATEMENT

### 2.1 System and Threat Model

We consider a computation outsourcing architecture involving two different entities, as illustrated in Fig. 1: the cloud customer, who has large amount of computationally expensive LP problems to be outsourced to the cloud; the cloud server (CS), which has significant computation resources and provides utility computing services, such as hosting the public LP solvers in a pay-per-use manner. The customer has a large-scale linear programming problem (to be formally defined later) to be solved. However, due to the lack of computing resources, like processing power, memory, and storage etc., he cannot carry out such expensive Computation locally. Thus, the customer resorts to CS for solving the LP computation and leverages its computation capacity in a pay-per-use manner. Instead of directly sending original problem, the customer first uses a secret K to map into some encrypted version K and outsources problem K to CS. CS then uses its public LP solver to get the answer of K and provides a correctness proof □, but it is supposed to learn nothing or little of the sensitive information contained in the original problem description. After receiving the solution of encrypted problem K, the customer should be able to first verify the answer via the appended proof . If it's correct, he then uses the secret K to map the output into the desired answer for the original problem.

The cloud user (U), who has large amount of data files to be stored in the cloud; the cloud server (CS), which is managed by cloud service provider (CSP) to provide data storage service and has significant storage space and computation resources (we will not differentiate CS and CSP hereafter.); the third party auditor (TPA), who has expertise and capabilities that cloud users do not have and is trusted to assess the cloud storage service security on behalf of the user upon request. Users rely on the CS for cloud data storage and maintenance. They may also dynamically interact with the CS to access and update their stored data for various application purposes.

The users may resort to TPA for ensuring the storage security of their outsourced data, while hoping to keep their data private from TPA. We consider the existence of a semi-trusted CS in the sense that in most of time it behaves properly and does not deviate from the prescribed protocol execution. While providing the cloud data storage based services, for their own benefits the CS might neglect to keep or deliberately delete rarely accessed data files which belong to ordinary cloud users. Moreover, the CS may decide to hide the data corruptions caused by server hacks or Byzantine failures to maintain reputation. We assume the TPA, who is in the business of auditing, is reliable and independent, and thus has no incentive to collude with either the CS or the users during the auditing process. TPA should be able to efficiently audit the cloud data storage without local copy of data and without bringing in additional on-line burden to cloud users. However, any possible leakage of user's outsourced data towards TPA through the auditing protocol should be prohibited.

Note that to achieve the audit delegation and authorize CS to respond to TPA's audits, the user can sign a certificate granting audit rights to the TPA's public key, and all audits from the TPA are authenticated against such a certificate. These authentication handshakes are omitted in the following presentation.

### 2.2 Design Goals

To enable privacy-preserving public auditing for cloud data storage under the aforementioned Model, our protocol design should achieve the following security and performance guarantee:

1) Public audit ability: to allow TPA to verify the correctness of the cloud data on demand without

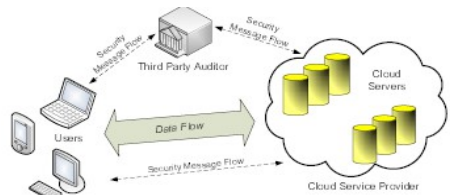

Fig 2: Architecture of cloud storage data service.

retrieving a copy of the whole data or introducing additional on-line burden to the cloud users;
2) Storage correctness: to ensure that there exists no cheating cloud server that can pass the audit from TPA without indeed storing users' data intact;
3) Privacy-preserving: to ensure that there exists no way for TPA to derive users' data content from the information collected during the auditing process;
4) Batch auditing: to enable TPA with secure and efficient auditing capability to cope with multiple auditing delegations from possibly large number of different users simultaneously; 5) Lightweight: to allow TPA to perform auditing with minimum communication and computation overhead.

### 2.3 Notation and Preliminaries

The data file to be outsourced, denoted as a sequence of n blocks $m_1, \ldots, m_n \in Z_p$ for some large prime p.
– $f_{key}(\cdot)$ – pseudorandom function (PRF), defined as: $\{0, 1\}* \times key \to Z_p$.
– $\_key(\cdot)$ – pseudorandom permutation (PRP), defined as: $\{0, 1\}^{\log_2(n)} \times key \to \{0,\}^{\log_2(n)}$.
– $MAC_{key}(.)$ – message authentication code (MAC) function, defined as: $\{0, 1\}* \times key \to \{0, 1\}^l$.
– $H(\cdot), h(\cdot)$ – map-to-point hash functions, defined as: $\{0, 1\}* \to G$, where G is some group.

We now introduce some necessary cryptographic background for our proposed scheme. Bilinear Map Let G1, G2 and GT be multiplicative cyclic groups of prime order p. Let g1 and g2 be generators of G1 and G2, respectively. A bilinear map is a map $e : G1 \times G2 \to GT$ with the following properties [15, 16]: 1) Computable: there exists an efficiently computable algorithm for computing e; 2) Bilinear: for all $u \in G1, v \in G2$ and $a, b \in Z_p$, $e(u^a, v^b) = e(u, v)^{ab}$; 3) Non-degenerate: $e(g1, g2) \neq 1$; 4) for any $u1, u2 \in G1, v \in G2, e(u1 u2, v) = e(u1, v)* e(u2, v)$.

## 3. PROPOSED SCHEMES

In the introduction we motivated the public audit ability with achieving economies of scale for cloud computing. This section presents our public auditing scheme for cloud data storage security. We start from the overview of our public auditing system and discuss two straightforward schemes and their demerits. Then we present our main result for privacy-preserving public auditing to achieve the aforementioned design goals. We also show how to extent our main scheme to support batch auditing for TPA upon delegations from multi-users. Finally, we discuss how to adapt our main result to support data dynamics.

### 3.1 Definitions and Framework of Public Auditing System

We follow the similar definition of previously proposed schemes in the context of remote data integrity checking and adapt the framework for our privacy-preserving public auditing system.

A public auditing scheme consists of four algorithms (KeyGen, SigGen, GenProof, VerifyProof). KeyGen is a key generation algorithm that is run by the user to setup the scheme. SigGen is used by the user to generate verification metadata, which may consist of MAC, signatures, or other related information that will be used for auditing. GenProof is run by the cloud server to generate a proof of data storage correctness, while VerifyProof is run by the TPA to audit the proof from the cloud server.

Our public auditing system can be constructed from the above auditing scheme in two phases,
**Setup and Audit:**
– Setup: The user initializes the public and secret parameters of the system by executing KeyGen, and pre-processes the data file F

by using SigGen to generate the verification metadata. The user then stores the data file F at the cloud server, deletes its local copy, and publishes the verification metadata to TPA for later audit. As part of pre-processing, the user may alter the data file F by expanding it or including additional metadata to be stored at server.

– Audit: The TPA issues an audit message or challenge to the cloud server to make sure that the cloud server has retained the data file F properly at the time of the audit. The cloud server will derive a response message from a function of the stored data file F by executing GenProof. Using the verification metadata, the TPA verifies the response via Verify Proof. Note that in our design, we do not assume any additional property on the data file, and thus regard error correcting codes as orthogonal to our system. If the user wants to have more error resiliency, he/she can first redundantly encode the data file and then provide us with the data file that has error-correcting codes integrated.

### A) Mechanism Design Framework

We propose to apply problem transformation for mechanism design. The general framework is adopted from a generic approach [9], while our instantiation is completely different and novel. In this framework, the process on cloud server can be represented by algorithm Proof Gen and the process on customer can be organized into three algorithms (KeyGen, ProbEnc, ResultDec).

These four algorithms are summarized below and will be instantiated later.

- KeyGen($1^k$) → {K}. This is a randomized key generation algorithm which takes a system security parameter k, and returns a secret key K that is used later by customer to encrypt the target LP problem.
- ProbEnc(K,_) → {_K}. This algorithm encrypts the input tuple into K with the secret key K. According to problem transformation, the encrypted input K and thus defines the problem to be solved in the cloud.
- ProofGen(_K) → {(y, □)}. This algorithm augments a generic solver that solves the problem K to produce both the output y and a proof . The output y later decrypts to x, and □ is used later by the customer to verify the correctness of y or x.
- ResultDec(K,_, y, □) → {x,⊥}. This algorithm may choose to verify either y or x via the proof. In any case, a correct output x is produced by decrypting y using the secret K.

The algorithm outputs when the validation fails, indicating the cloud server was not performing the computation faithfully. Note that our proposed mechanism provides us one-time pad types of flexibility. Namely, we shall never use the same secret key K to two different problems. Thus, when analyzing the security strength of the mechanism, we focus on the cipher text only attack. We do not consider known plaintext attack in this paper but do allow adversaries to do offline guessing or inferring via various problem-dependent information including sizes and signs of the solution, which are not necessary to be confidential.

### B) Basic Techniques

Before presenting the details of our proposed mechanism, we study in this subsection a few basic techniques and show that the input encryption based on these techniques along may result in an unsatisfactory mechanism. However, the analysis will give insights on how a stronger mechanism should be designed. Note that to simplify the presentation, we assume that the cloud server honestly performs the computation, and defer the discussion on soundness to a later section.

1) Hiding equality constraints (A, b): First of all, a randomly generated m × m non-singular matrix Q can be part of the secret key K. The customer can apply the matrix to Eq. (2) for the following constraints transformation, $Ax = b \Rightarrow A'x = b'$ where $A' = QA$ and $b' = Qb$.

Since we have assumed that A has full row rank, A′ must have full row rank. Without knowing Q, it is not possible for one to determine the exact elements of A. However, the Null spaces of A and A′ remains the same, which may violate the security requirement of some applications. The vector b is encrypted in a perfect way since it can be mapped to an arbitrary b′ with a proper choice of Q.

2) Hiding inequality constraints (B): The customer cannot transform the inequality constraints in the similar way as used for the equality constraints. This is because for an arbitrary invertible matrix Q, $Bx \geq 0$ is not equivalent to $QBx \geq 0$ in general.

To hide B, we can leverage the fact that a feasible solution to Eq. (2) must satisfy the equality constraints. To be more specific, the feasible regions defined by the following two groups of constraints are the same.

$Ax = b$
$Bx \geq 0$
$Ax = b$
$(B - \_A)x = B'x \geq 0$

where A is a randomly generated n×m matrix in K satisfying that $|B'| = |B - \_A| \neq 0$ and $\_b = 0$. Since the condition b = 0 is largely underdetermined, it leaves great flexibility to choose _ in order to satisfy the above conditions.

3) Hiding objective functions c and value $c^T x$: Given the widely application of LP, such as the estimation of business annul revenues or personal portfolio holdings etc., the information contained in objective function c and optimal objective value $c^T x$ might be as sensitive as the constraints of A,B, b. Thus, they should be protected, too.

To achieve this, we apply constant scaling to the objective function, i.e. a real positive scalar is generated randomly as part of encryption key K and c is replaced by c. It is not possible to derive the original optimal objective value $c^T x$ without knowing first, since it can be mapped to any value with the same sign. While hiding the objective value well, this approach does leak structure-wise information of objective function c. namely; the number and position of zero-elements in c are not protected. Besides, the ratio between the elements in c are also preserved after constant scaling.

**Summarization of basic techniques** Overall, the basic techniques would choose a secret key K = (Q) and encrypt the input tuple into K = (A′,B′, b′, c), which gives reasonable strength of problem input hiding. Also, these techniques are clearly correct in the sense that solving K would give the same optimal solution as solving. However, it also implies that although input privacy is achieved, there is no output privacy. Essentially, it shows that although one can change the constraints to a completely different form, it is not necessary the feasible region defined by the constraints will change, and the adversary can leverage such information to gain knowledge of the original LP problem. Therefore, any secure linear programming mechanism must be able to not only encrypt the constraints but also to encrypt the feasible region defined by the constraints.

# 4. SECURITY ANALYSIS

## 4.1 Security proof

We evaluate the security of the proposed scheme by analyzing its fulfillment of the security guarantee described in Section 2, namely, the storage correctness and privacy-preserving. We start from the single user case, where our main result is originated. Then we show how to extend the, security guarantee to a multi-user setting, where batch auditing for TPA is enabled. All proofs are derived on the probabilistic base, i.e., with high probability assurance, which we omit writing explicitly. Storage Correctness Guarantee We need to prove that the cloud server can not generate valid response toward TPA without faithfully storing the data, as captured by Theorem1.

*Theorem 1*. If the cloud server passes the Audit phase, then it must indeed possess the specified
data intact as it is.

Proof (Proof Sketch). The proof consists of three steps. First, we show that there exists no ma-licious server that can forge a valid response $\{\_, \mu, R\}$ to pass the verification equation 1. The correctness of this statement follows from the Theorem 4.2 proposed in [11]. Note that the value R in our protocol, which enables the privacy-preserving guarantee, will not affect the validity of the equation, due to the circular relationship between R and in $= h(R)$ and the verification equation. Next, we show that if the response $\{\_, \mu, R\}$ is valid, where $\mu = \mu' + r$ and $= h(R) = h(e(u, v)r)$, then the underlying $\mu'$ must be valid too. Indeed, we can extract $\mu'$ from the protocol in the random oracle model. Finally, similar to the argument in [11], we show that the validity of $\mu'$ implies the correctness of $\{m_i\}_{i \in I}$ where $\mu' = P_{i \in I} \_i m_i$. Here we utilize the small exponent (SE) test technique of batch verification in [21]. Because $\{\_i\}$ are picked up randomly by the TPA in each Audit phase, $\{\_i\}$ can be viewed similarly as the random chosen exponents in the SE test [21]. Therefore, the correctness of individual sampled blocks is ensured. All above sums up to the storage correctness guarantee. Privacy Preserving Guarantee We want to make sure that TPA can not derive users' data content from the information collected during auditing process. This is equivalent to prove the Theorem 2. Note that if $\mu'$ can be derived by TPA, then $\{m_i\}_{i \in I}$ can be easily obtained by solving a group of linear equations when enough combinations of the same blocks are collected.

Theorem 2. From the server's response $\{\_, \mu, R\}$, TPA cannot recover $\mu'$.

Proof (Proof Sketch). Again, we argue in three steps. First, recall the relationship between $\mu'$ and $\mu'$, which requires solving discrete-log problems.

Second, we consider how to learn $\mu'$ from $\mu$. Note that $\mu$ is blinded by r as $\mu = \mu' + r$ and $R = e(u, v)r$, where r is chosen randomly by cloud server and is unknown to TPA. Even with R, due to the hardness of discrete-log assumption, the value r is still hidden against TPA. Thus, privacy of $\mu'$ is guaranteed from $\mu$.

Finally, all that remains is to prove from $\{\_, \mu, R\}$, still no information on $\mu'$ can be obtained by TPA. Recall that r is a random private value chosen by the server and $\mu = \mu' + r$, where $= h(e(u, v)r)$. Following the same technique of Schnorr signature, our auditing protocol between TPA and cloud server can be regarded as a provably secure honest zero knowledge identification scheme, by viewing $\mu'$ as a secret key and as a challenge value, which implies no information
on $\mu'$ can be leaked. Indeed, it is easy to simulate valid response $\{\mu, R\}$ without knowing $\mu'$ in the random oracle model. This completes the proof of Theorem 2.

Security Guarantee for Batch Auditing Now we show that extending our main result to a multi-user setting will not affect the aforementioned security insurance, as shown in Theorem 3:

Theorem 3. Our batch auditing protocol achieves the same storage correctness and privacy pre- serving guarantee as in the single-user case.

Proof (Proof Sketch). We only prove the storage correctness guarantee, as the privacy-preserving guarantee in the multi-user setting is similar to that of Theorem 2, and thus omitted here. The proposed batch auditing protocol is built upon the aggregate signature scheme proposed in [15].

According to the security strength of aggregate signature [15], in our multi-user setting, there exists no malicious cloud servers that can forge valid $\mu_1, \ldots, \mu_k$ in the responses to pass the verification equation 2. Actually, the equation 2 functions as a kind of screening test as proposed.

While the screening test may not guarantee the validity of each individual $\_k$, it does ensure the authenticity of $\mu_k$ in the batch auditing protocol, which is adequate for the rationale in our case.

Once the validity of $\mu_1, \ldots, \mu_k$ is guaranteed, from the proof of Theorem 1, the storage correctness guarantee in the multi-user setting is achieved.

## 4.2 Performance Analysis

We now assess the performance of the proposed privacy-preserving public auditing scheme. We will focus on the extra cost introduced by the privacy-preserving guarantee and the efficiency of the proposed batch auditing technique. The experiment is conducted using C on a Linux system with an Intel Core 2 processor running at 1.86 GHz, 2048MB of RAM, and a 7200 RPM Western Digital 250 GB Serial ATA drive with an 8 MB buffer. Algorithms use the Pairing-Based Cryptography (PBC) library version 0.4.18. The elliptic curve utilized in the experiment is a MNT curve, with base field size of 159 bits and the embedding degree 6. The security level is chosen to be 80 bit, which means $|i| = 80$ and $|p| = 160$. All experimental results represent the mean of 20 trials.

Cost of Privacy-preserving Guarantee We begin by estimating the cost in terms of basic cryptographic operations, as notated in Table 1. Suppose there are c random blocks specified in the chal during the Audit phase. Under this setting, we quantify the extra cost introduced by the support of privacy-preserving into server computation, auditor computation as well as communication overhead. On the server side, the generated response includes an aggregated signature $A = Q_i *I I_i, G_1$, a random metadata $R = e(u, v)r \in G_T$, and a blinded linear combination of sampled blocks $\mu = P_{i \in I} \_i m_i + r \in Z_p$, where $= h(R) \in Z_p$. The corresponding computation cost is c-MultExp1
$G_1(|\_i|)$, Exp1
$G_T (|p|)$, and Hash1
$Z_p + Add_c$
$Z_p + Mult_{c+1}$
$Z_p$, respectively.

Compared to the existing homomorphic authenticator based solution for ensuring remote data integrity, the extra cost for protecting the user privacy, resulted from the random mask R, is only a constant:
Exp1
$G_T (|p|) + Mult_1$
$Z_p + Hash_1$

Zp + Add1
Zp , which has nothing to do with the number of sampled
blocks c. When c is set to be 460 or 300 for high assurance of auditing, as discussed in Section 3.3, the extra cost for privacy-preserving guarantee on the server side would be negligible against the total server computation for response generation.
Similarly, on the auditor side, upon receiving the response {_,R, μ}, the corresponding computation cost for response validation is Hash1
Zp+c-MultExp1
G1(|_i|)+Hashc
G1+Mult1
G1+Mult1
GT +
Exp3
G1 (|p|) + Pair2
G1,G2 , among which only Hash1
Zp + Exp2
G1(|p|) + Mult1
GT account for the additional constant computation cost. For c = 460 or 300, and considering the relatively expensive pairing operations, this extra cost imposes little overhead on the overall cost of response validation, and thus can be ignored. For the sake of completeness, Table 2 gives the experiment result on performance comparison between our scheme and the state-of-the-art. It can be shown that the performance of our scheme is almost the same as that of [11], even if our scheme supports privacy-preserving guarantee while [11] does not. Note that in our scheme, the server's response {,R, μ} contains an additional random element R, which is a group element of GT and has the size close to 960 bits. This explains the extra communication cost of our scheme opposing. Batch Auditing Efficiency Discussion in Section 3.4 gives an asymptotic efficiency analysis on the batch auditing, by considering only total number of expensive pairing operations. However, on the practical side, there are additional operations required for batching, such as modular exponentiations and multiplications. Meanwhile, the different sampling strategies, i.e., different numbers.

## 5. RELATED WORK

General secure computation outsourcing that fulfills all aforementioned requirements, such as input/output privacy and correctness/soundness guarantee has been shown feasible in theory by Gennaro et al. [9]. However, it is currently not practical due to its huge computation complexity. Instead of outsourcing general functions, in the security community, Atallah et al. explore a list of work [5], [7], [8], [10] for securely outsourcing specific applications. The customized solutions are expected to be more efficient than the general way of constructing the circuits. In [5], they give the first investigation of secure outsourcing of numerical and scientific computation. A set of problem dependent disguising techniques are proposed for different scientific applications like linear algebra, sorting, string pattern matching, etc. However,
these disguise techniques explicitly allow information disclosure to certain degree. Besides, they do not handle the important case of result verification, which in our work is bundled into the design and comes at close-to-zero additional cost. Later on in [7] and [8], Atallah et al. give two protocol designs for both secure sequence comparison outsourcing and secure algebraic computation outsourcing. However, both protocols use heavy cryptographic primitive such as homomorphic encryptions and/or oblivious transfer and do not scale well for large problem set. In addition, both designs are built upon the assumption of two non-colluding servers and thus vulnerable to colluding attacks. Based on the same assumption, Hohenberger et al. [6] provide protocols for secure outsourcing of modular exponentiation, which is considered as prohibitively expensive in most public-key cryptography operations. Very recently, Atallah [10] et al. give a provably secure protocol for secure outsourcing matrix multiplications based on secret sharing While this work outperforms their previous work [8] in the sense of single server assumption and computation efficiency (no expensive cryptographic primitives), the drawback is the large communication overhead. Namely, due to secret sharing technique, all scalar operations in original matrix multiplication are expanded to polynomials, introducing significant amount of overhead.
Considering the case of the result verification, the communication overhead must be further doubled, due to the introducing of additional pre-computed "random noise" matrices.
In short, these solutions, although elegant, are still not efficient enough for immediate practical uses, which we aim to address for the secure LP outsourcing in this paper. B. Work on Secure Multiparty Computation Another large existing list of work that relates to (but is also significantly different from) ours is Secure Multi-party Computation (SMC), first introduced by Yao [11] and later extended by Goldreich and many others. SMC allows two or more parties to jointly compute some general function while hiding their inputs to each other. As general SMC can be very inefficient, Du and Atallah et. al. have proposed a series of customized solutions under the SMC context to a spectrum of special computation problems, such as privacy-preserving cooperative statistical analysis, scientific computation, geometric computations, sequence comparisons, etc. However, directly applying these approaches to the cloud computing model for secure computation outsourcing would still be problematic. The major reason is that they did not address the asymmetry among the computational powers possessed by cloud and the customers, i.e., all these schemes in the context of SMC impose each involved parties comparable computation burdens, which we specifically avoid in the mechanism design by shifting as much as possible computation burden to cloud only. Another reason is the asymmetric security requirement. In SMC no single involved party knows all the problem input information, making result verification a very difficult task. But in our model, we can explicitly exploit the fact that the customer knows all input information and thus design efficient result verification mechanism.
Recently, Li and Atallah give a study for secure and collaborative computation of linear programming under the SMC framework. Their solution is based on the additive split of the constraint matrix between two involved parties, followed by a series of interactive (and arguably heavy) cryptographic protocols collaboratively executed in each iteration step of the Simplex Algorithm. This solution has the computation asymmetry issue mentioned previously. Besides, they only consider honest-but-curious model and thus do not guarantee that the final solution is optimal.

### A. Work on Delegating Computation and Cheating Detection

Detecting the unfaithful behaviors for computation outsourcing is not an easy task, even without consideration of input/output privacy. Verifiable computation delegation, where a computationally weak customer can verify the correctness of the delegated computation results from a powerful but untrusted server without investing too much resources, has found great interests in theoretical computer science community. Some recent general result can be found in Goldwasser et al.. In distributed computing and targeting the specific computation delegation of one-way function inversion, Golle et al propose to insert some pre-computed results (images of "ringers") along with the computation workload to defeat untrusted (or lazy) workers. Du. et al. propose a method of cheating detection for general computation outsourcing in grid computing. The server is required to provide a commitment via a Merkle tree based on the results it computed.

The customer can then use the commitment combined with a sampling approach to carry out the result verification (without re-doing much of the outsourced work.). However, all above schemes allow server actually see the data and result it is computing with, which is strictly prohibited in the cloud computing model for data privacy. Thus, the problem of result verification essentially becomes more difficult, when both input/output privacy is demanded. Our work leverages the duality theory of LP problem and effectively bundles the result verification within the mechanism design, with little extra overhead on both customer and cloud server.

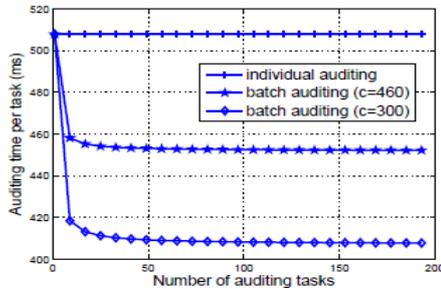

Fig 3: Comparison on auditing time between batch auditing and individual auditing. Per task auditing time denotes the total auditing time divided by the number of tasks. For clarity reasons, we omit the straight curve for individual auditing when c=300.

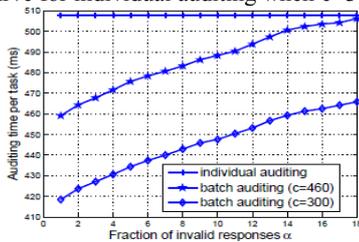

Fig 4: Comparison on auditing time between batch auditing and individual auditing, when α-fraction of 256 responses are invalid. Per task auditing time denotes the total auditing time divided by the number of tasks.

of sampled blocks c, is also a variable factor that affects the batching efficiency. Thus, whether the benefits of removing pairings significantly outweighs these additional operations is remained to be verified. To get a complete view of batching efficiency, we conduct a similar timed batch auditing test as , where the number of auditing tasks is increased from 1 to approximately 200 with intervals of 8. The performance of the corresponding non-batched (individual) auditing is provided as a baseline for the measurement. Following the same experimental setting as c = 460 and 300, the average per task auditing time for both batch auditing and the individual auditing is shown in Fig. 2, where the per task auditing time is computed by dividing total auditing time by the number of tasks. It can be shown that compared to individual auditing, batch auditing indeed helps reduce the TPA's computation cost, as more than 11% and 14% of per-task auditing time is saved, when c is set to be 460 and 300, respectively.

Sorting out Invalid Responses Now we use experiment to justify the efficiency of our recursive binary search approach for TPA to sort out the invalid responses when batch auditing fails, as Note that this experiment is tightly pertained to works by [11, 13] which evaluates the batch verification efficiency of various short signature schemes.

To evaluate the feasibility of the recursive approach, we first generate a collection of 256 valid responses, which implies the TPA may concurrently handle 256 different auditing delegations. We then conduct the tests repeatedly while randomly corrupting an fraction, ranging from 0 to 18%, by replacing them with random values. The average auditing time per task against the individual auditing approach is presented in Fig. 3. The result shows that even the number of invalid responses exceeds 15% of the total batch size, the performance of batch auditing can still be safely concluded as more preferable than the straightforward individual auditing.

Note that this is consistent with the experiment results derived.

## 6. CONCLUSION

In this paper, we propose a privacy-preserving public auditing system for data storage security in Cloud Computing, where TPA can perform the storage auditing without demanding the local copy of data. We utilize the homomorphic authenticator and random mask technique to guarantee that TPA would not learn any knowledge about the data content stored on the cloud server during the efficient auditing process, which not only eliminates the burden of cloud user from the tedious and possibly expensive auditing task, but also alleviates the users' fear of their outsourced data leakage along with the problem of securely outsourcing LP computations in cloud computing, and provide such a practical mechanism design which fulfills input/output privacy, cheating resilience, and efficiency. By explicitly decomposing LP computation outsourcing into public LP solvers and private data, our mechanism design is able to explore appropriate security/efficiency tradeoffs via higher level LP computation than the general circuit Representation.

## Refrences

## Author's Profile

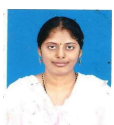

**Mrs G Vidhisha** received her B.Tech in Bioinformatics from MITS-Madanapalli and now pursuing M.Tech (CSE) Vaagdevi Institute of Technology and Sciences, JNTU-Anantapur.

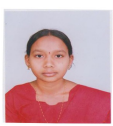

**Miss C Surekha** received her B.Tech in CSE from VITS-PDTR and now pursuing M.Tech (CSE) Vaagdevi Institute of Technology and Sciences, JNTU-Anantapur.

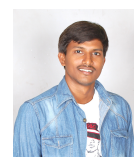

**Mr S Sanjeeva Rayudu** received his B Tech in CSE from VITS-Pdtr, and his M.Tech in CSE from VITS-PDTR, JNTUA. Working as a Asst Prof in CSE in Vaagdevi Institute of Technology and Sciences, JNTU-Anantapur.

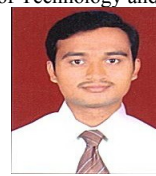

**Mr. U.Seshadri** received his M.Sc (Mathematics) from Sri Venkateswara University-Tirupati. M.Tech in Computer Science and Engineering from Satyabhama University. And working as a HOD in Computer Science and Engineering,MCA in Vaagdevi Institute of Technology and Sciences, JNTU-Anantapur.